\documentclass{article}
\usepackage{arxiv}

\usepackage[utf8]{inputenc}
\usepackage[T1]{fontenc}
\usepackage{hyperref}
\usepackage{url}
\usepackage{booktabs}
\usepackage{amsfonts}
\usepackage{nicefrac}
\usepackage{microtype}
\usepackage{graphicx}
\usepackage[numbers,sort&compress]{natbib}
\usepackage{subcaption}
\usepackage{xcolor}
\usepackage{doi}
\usepackage{multirow}
\usepackage{CJKutf8}

\usepackage{longtable,booktabs,array}
\usepackage{pdflscape}
\usepackage{wasysym}
\makeatletter
\renewcommand{\section}{\@startsection{section}{1}{\z@}{-1.4ex \@plus -0.3ex \@minus -0.2ex}{0.7ex \@plus 0.2ex}{\Large\bf\raggedright}}
\renewcommand{\subsection}{\@startsection{subsection}{2}{\z@}{-1.1ex \@plus -0.3ex \@minus -0.2ex}{0.4ex \@plus 0.1ex}{\large\bf\raggedright}}
\renewcommand{\subsubsection}{\@startsection{subsubsection}{3}{\z@}{-1.0ex \@plus -0.3ex \@minus -0.2ex}{0.2ex \@plus 0.1ex}{\normalsize\bf\raggedright}}
\makeatother

\title{Geospatial Dynamics of Online Attention to Nuclear Power in Japan Before and After Fukushima}

\date{}

\author{Yifan Sun\thanks{Corresponding author: \texttt{sun.yifan.7r@kyoto-u.ac.jp}} \\
	Institute for Integrated Radiation and Nuclear Science, Kyoto University \\
	Kumatori, Osaka 590-0494, Japan \\
    \And
	Ken Kurosaki \\
	Institute for Integrated Radiation and Nuclear Science, Kyoto University \\
	Kumatori, Osaka 590-0494, Japan \\
    \And
	Benjamin C. McLellan \\
	Graduate School of Energy Science, Kyoto University \\
	Yoshida-Honmachi, Sakyo-ku, Kyoto 606-8501, Japan \\
}

\date{}

\renewcommand{\headeright}{}
\renewcommand{\undertitle}{}
\renewcommand{\shorttitle}{}

\begin{document}
\maketitle

\begin{abstract}
Nuclear power is increasingly seen as necessary for sustainable development, but this role depends on informed public engagement and support. Building this support requires understanding what citizens want to know, as reflected in their spontaneous online searches. To map this demand, we present the first long-term (2004–2024), prefecture-level analysis of Google Trends data on nuclear power in Japan. The results indicate that in routine periods, search interest followed a host-locality pattern, being significantly higher in prefectures hosting nuclear plants, with little spillover to their neighbors. Immediately after the Fukushima Daiichi accident in March 2011, this pattern changed, and interest instead declined with distance from Fukushima, before the host-locality pattern returned from 2012. During March 2011, searches leaned heavily toward general and live information rather than guidance on what to do. Their framing also varied with proximity, as those near Fukushima prioritized live updates and formal countermeasures, while others leaned toward background information and personal-scale questions. These patterns and the shift between them provide a behavioral evidence base for designing communication strategies in Japan and countries weighing the use of nuclear power. More broadly, the finding that searches vary with proximity can inform public communication in other geographically anchored crises.
\end{abstract}

\section{Introduction}

Following the Fukushima Daiichi (1F) accident of March 2011, Japan's nuclear fleet underwent a complete shutdown. Before the accident, nuclear power had supplied close to 30\% of Japan's electricity, and was also a key element of government energy policy, so the shutdown increased the country's dependence on imported fossil fuels and raised its fuel costs, posing a major challenge for Japan's energy security, economy, and environment \citep{vivoda2012japan}. The accident also affected nuclear policy and public opinion beyond Japan. One of the most notable examples is Germany, which quickly shut down its oldest reactors, a response that has been attributed to intense domestic media coverage, electoral timing, and felt cultural proximity to Japan \citep{wittneben2012impact}. \citet{kim2013effect} observed an overall decrease in public acceptance of nuclear energy by analyzing nearly 25,000 poll respondents across 42 countries, with the decrease correlating with factors such as existing nuclear infrastructure, government control of the media, and each country's distance from Fukushima. As policymakers on multiple continents now revisit nuclear's role in low-carbon transitions, post-Fukushima nuclear governance, involving restart decisions, new construction, and waste management, depends on regaining public trust, which requires understanding what the public thinks and wants to know.

Surveys and polls are the conventional way to ask the public what they think about nuclear power, and they have produced a substantial body of evidence on attitudes across decades and across countries. For example, studies in the UK \citep{pidgeon2008climate,bickerstaff2008reframing} found that public opinion on nuclear power is not fixed but depends on how it is framed. Acceptance rose when nuclear power was presented as a tool for combating climate change, but it amounted to a "reluctant acceptance" rather than active support. A similar framing effect has been observed for new construction in Japan, where respondents, about 40\% of whom were neutral toward new plants, became more supportive when the plants were described as no longer requiring evacuation, but not when a disposal site for radioactive waste was assumed to be selected \citep{yamagata2024public}. 
Opinion also varies with respondents' own circumstances. For instance, men and those with more education were more supportive of nuclear power, as were those with higher and less elastic electricity demand \citep{arikawa2014attitudes}. Households on average wanted far more renewables in the electricity mix than Japan actually used (in 2017), and those living near nuclear plants operated by the Tokyo Electric Power Company favored an even higher share \citep{okubo2020preferences}. Looking across decades of Japanese polls, \citet{kitada2016public} found that clear opposition to nuclear power utilization, a minority view of 20--30\% before the 1F accident, rose to 70\% months afterward. Polls by Kitada's own institute also suggested that people found the consequences of reducing nuclear power hard to grasp. While few thought a shift to renewable energy feasible within 10 years, only about a third expected CO$_2$ emissions to rise if nuclear power was reduced, and few were willing to accept a large rise in electricity bills to pay for this shift.

While surveys provide structured, systematic measurement, they have known limitations. On sensitive topics, where some answers are more socially acceptable than others, respondents often misreport, by deliberately or subconsciously editing their answers to avoid embarrassment in front of an interviewer or repercussions from third parties \citep{tourangeau2007sensitive}. Polarized issues such as nuclear power may invite the same bias. More importantly, surveys may struggle to capture the in-the-moment information demand of an unfolding crisis as even rapid post-disaster needs assessments generally begin days after the event and can take days to complete \citep{korteweg2010rapid}. Voluntary search data, on the other hand, offer a complementary signal to surveys. Searches are made without a third-party interviewer and at the moment of need, so they reveal actual (rather than stated) preferences and opinions, and are largely free from recall and social desirability biases \citep{mavragani2019predicting,eysenbach2009infodemiology,brodeur2021covid}. Because they are recorded continuously, search data are also resilient to unexpected events and remain usable when circumstances make surveys difficult to conduct, as during the COVID-19 lockdowns \citep{brodeur2021covid}.

\citet{lazer2009computational} made the case for a computational social science that collects and analyzes data at an unprecedented scale, and pointed to Internet searches as a valuable source of such data. Since then, a growing body of work across many fields has used Google Trends, a free and easy-to-use tool built on billions of daily searches \citep{jun2018ten}. For example, \citet{ginsberg2009detecting} demonstrated that aggregated Google queries could estimate weekly influenza-like-illness rates one to two weeks ahead of official US reports based on health-provider data; \citet{mellon2013and,mellon2014internet} validated Google Trends against traditional polling in the United States, United Kingdom, and Spain, demonstrating that carefully chosen search terms can track survey-measured public salience of issues; and \citet{choi2012predicting} demonstrated how Google Trends data can be used to improve short-term forecasts of economic indicators. Despite this wide application, search data have rarely been used to study public attention to nuclear power. The few studies in the nuclear or radiation domain are limited to illustrating the international public reaction to the Fukushima 1F accident \citep{zhu2016fear} and qualitatively describing worldwide searches for the single English term "radiation" \citep{mohamad2019using}. 

In Japan, this search signal remains underutilized: to our knowledge, no systematic, data-driven analysis has yet examined what the public has sought to know about nuclear power, or how this has changed over time. Such evidence would help governments, host communities, and operators match the information they provide to what the public actually seeks, as Japan restarts its reactors. To fill this gap, we present the first long-term, prefecture-level analysis of Google Trends data on nuclear power, covering two decades (2004--2024) that span routine operation, the 1F accident, and the subsequent shutdown and restarts. We address three research questions:
\begin{itemize}
    \item RQ1: What is the routine spatial distribution of online attention to nuclear power across Japan's prefectures?
    \item RQ2: How did the 1F accident reshape this distribution?
    \item RQ3: What did the public search for in the weeks after the 1F accident, and did this vary by place?
\end{itemize} 

By linking these data to each prefecture's reactor status, our analysis complements self-report surveys and polls by revealing what the public actively seeks to understand. Because Google Trends provides a consistent, continuous measure for every prefecture across two decades, we can also trace how this demand shifts through space and time, at a scale and resolution that periodic polls rarely reach. More broadly, our findings can inform public engagement in countries weighing the use and expansion of nuclear power and, because the 1F accident was also a geographically anchored crisis, can guide how to meet public information needs in other such crises.

\section{Results}
\subsection{Geospatial and Temporal Dynamics of Public Attention to Nuclear Power}

\subsubsection{Routine period (2004--2010): Localized attention}
\begin{figure}[htbp]
    \centering
    \includegraphics[width=\linewidth]{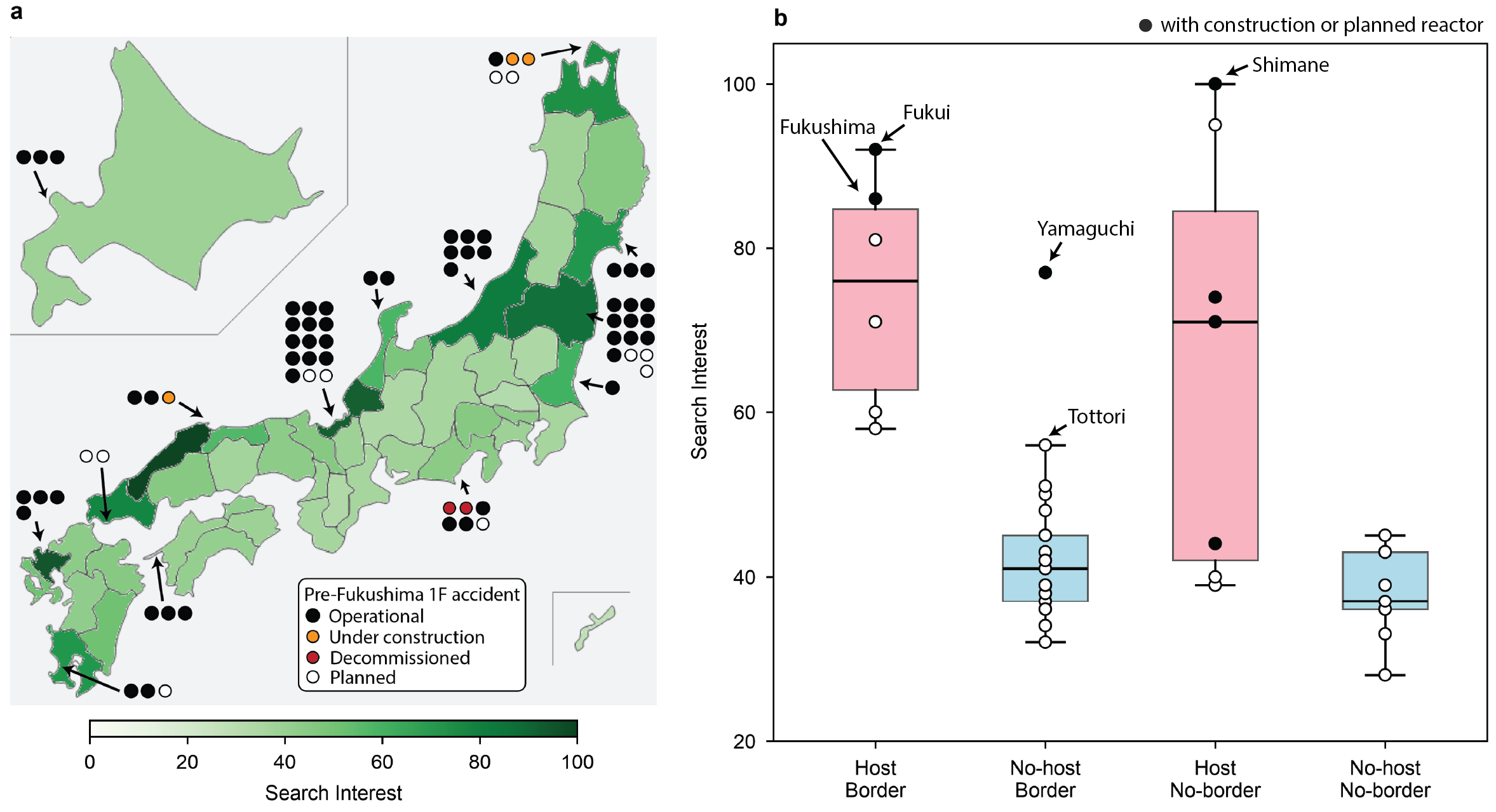}  %
    {\phantomsubcaption\label{fig:pre_fukushima_interest}}
    {\phantomsubcaption\label{fig:violin_pre_fukushima}}
    \caption{\textbf{Search Interest on the topic of "nuclear power plant" before the Fukushima 1F accident (2004--2010).} (a) Prefecture-level Search Interest from January 2004 to December 2010. Orange circles mark reactors under construction (Shimane-3, Ohma, TEPCO Higashidori-1), and red circles mark Hamaoka 1\&2, which ceased operation in 2009; 11 more reactors were planned before the Fukushima 1F accident. (b) Search Interest by host status ("Host/No-host", whether a prefecture itself hosts an operational reactor) and border status ("Border/No-border", whether it shares a land border with another prefecture hosting a reactor). Boxes show the median and interquartile range, with whiskers reaching the furthest prefecture within 1.5 times the interquartile range beyond the box; points are prefectures.}
    \label{fig:routine}
\end{figure}

Figure~\ref{fig:pre_fukushima_interest} maps prefecture-level Search Interest on the topic of nuclear power plants from January 2004 to December 2010, together with each prefecture's reactor status. Before the Fukushima 1F accident, attention formed a distinct multi-nodal pattern, concentrated in prefectures hosting operational reactors such as Shimane, Fukui, and Fukushima. To quantify this pattern, we further investigated whether prefectures hosting an operational reactor show higher search interest than non-hosts, and whether this elevated attention spills over into bordering prefectures that have no reactor of their own.

We first divided the 47 prefectures into four groups by host status (whether the prefecture itself has an operating reactor) and border status (whether it borders a host prefecture), and compared their search interest distributions (Figure~\ref{fig:violin_pre_fukushima}). Within each border group, two-sided Mann--Whitney U tests showed that host prefectures had significantly higher search interest than non-hosts (bordering, $p < 0.001$; non-bordering, $p = 0.013$; Supplementary Information, Table S1). Among host prefectures, those with reactors under construction or in planning, such as Shimane, Fukui, and Fukushima, show some of the highest search interest in the country, possibly because the progress of these projects and the debates around them keep them in the news. Such an effect is hard to confirm because the sample is small and because Fukui and Fukushima also hosted the most operating reactors, which confounds the comparison. Yamaguchi, however, is a notable outlier: with only the planned Kaminoseki Units 1 and 2 and no reactor operating or under construction, it has by far the highest search interest among prefectures without an operating reactor, even above the average of host prefectures. This suggests that becoming a prospective host can itself draw substantial attention, as residents search in anticipation to learn what hosting a plant would mean for them, even at the planning stage.

Border status alone, in contrast, had no statistically significant effect on search interest, either among hosts ($p = 0.668$) or among non-hosts ($p = 0.142$; Supplementary Information, Table S1). In general, then, the elevated attention in host prefectures did not spill over into neighboring prefectures without reactors. One case is still worth noting: Tottori, which borders Shimane but has no reactor of its own, has the second-highest search interest among non-host prefectures, after Yamaguchi. Tottori's tie to Shimane goes beyond a shared border in that some of its western cities lie within 30 km of the Shimane plant, an area that later, in 2012, became part of the plant's Urgent Protective Action Planning Zone. Even before then, one of these cities, Sakaiminato, had repeatedly asked the plant's operator for a safety agreement from 2005 \citep{sakaiminato_agreement}. In this case, the local stake and engagement that came with physical proximity to the plant, rather than prefectural adjacency alone, may explain the increased attention.

Before the Fukushima 1F accident, therefore, search interest followed host status, forming what we call the "host-locality" attention pattern: prefectures with operating or planned reactors showed elevated interest that, in general, did not spill over significantly into their neighbors.

\subsubsection{Fukushima 1F accident period (2011): Proximity-driven attention}
\begin{figure}[htbp]
    \centering
    \includegraphics[width=0.95\linewidth]{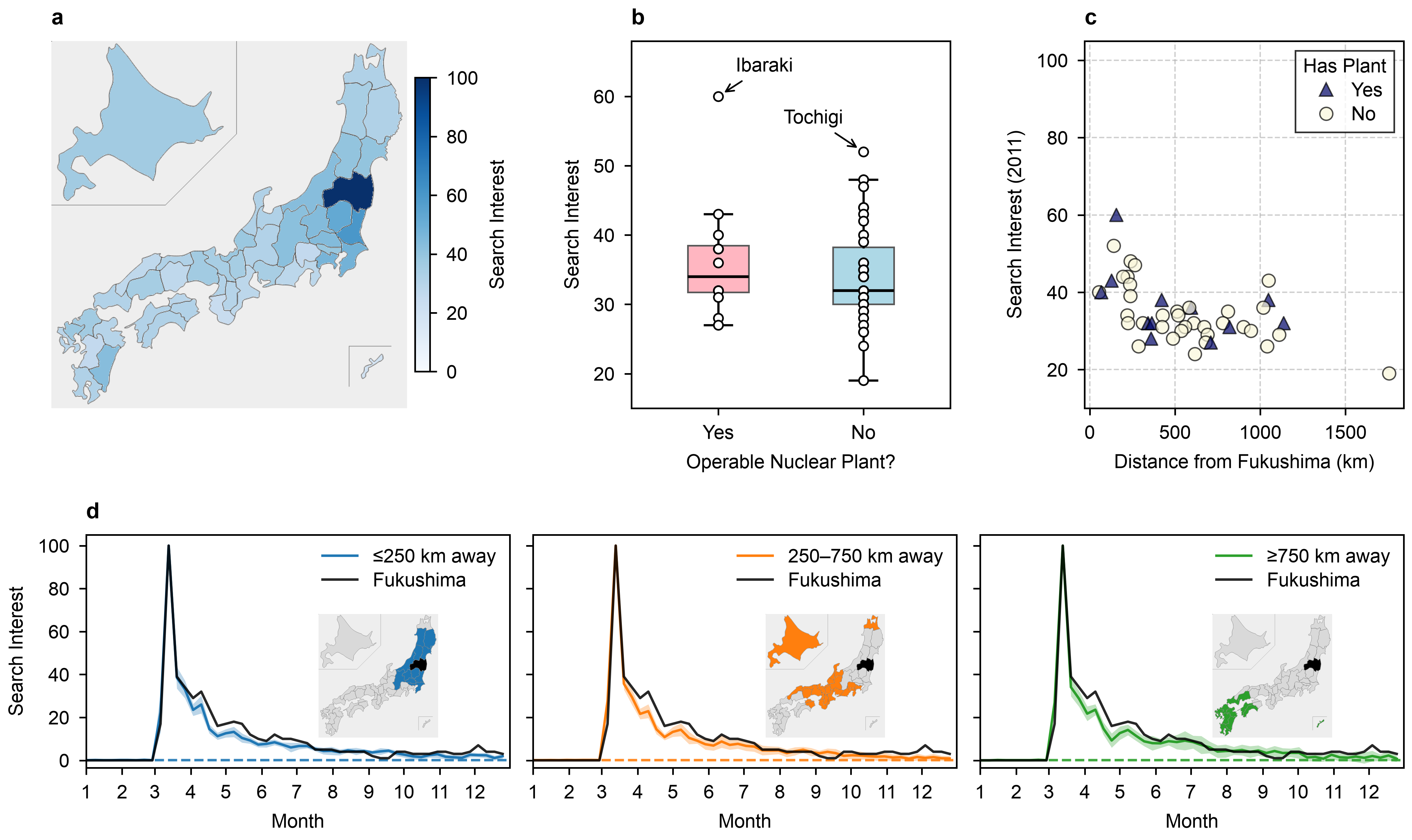}
    {\phantomsubcaption\label{fig:interest-map-2011}}
    {\phantomsubcaption\label{fig:during-fukushima-violin}}
    {\phantomsubcaption\label{fig:during-fukushima-distance-2011}}
    {\phantomsubcaption\label{fig:decay-rate}}
    \caption{\textbf{Search Interest in 2011.} (a) Prefecture-level spatial distribution. (b) Search Interest in prefectures with and without an operable reactor (one not yet retired) as of March 2011, excluding Fukushima Prefecture, which had the maximum of 100. (c) Search Interest against distance from Fukushima Prefecture. Fukushima itself is omitted because its distance to itself is zero. (d) Weekly Search Interest during 2011 for prefectures less than 250 km, 250 to 750 km, and over 750 km from Fukushima Prefecture (locations shown in the insets). Colored lines are group means with standard deviations shaded, and dashed lines show the January--February means.}
    \label{fig:during-fukushima}
\end{figure}

In 2011, in the immediate aftermath of the 1F accident, in place of the multi-nodal pattern of 2004--2010, attention concentrated in Fukushima Prefecture (Figure~\ref{fig:interest-map-2011}), whose search interest far exceeded that of every other prefecture. The effect of host status on attention was also disrupted, and prefectures with and without operable reactors no longer differed significantly (Mann--Whitney U test, $p = 0.475$). In contrast, elevated interest in Tochigi and Ibaraki, both bordering Fukushima, pointed to distance from the accident site as the new driver (Figure~\ref{fig:during-fukushima-violin}), so each prefecture's search interest was compared with its distance from Fukushima Prefecture. Before the accident (2004--2010), the two showed no significant correlation (Spearman's $\rho = 0.17$, $p = 0.253$; Supplementary Information, Figure S1a), and host prefectures generally showed higher attention regardless of distance. The turning point came in 2011, when search interest decreased significantly with distance (Spearman's $\rho = -0.54$, $p < 0.001$; Figure~\ref{fig:during-fukushima-distance-2011}), which we call the "proximity-driven" attention pattern. Okinawa, the farthest prefecture, consistently had the lowest interest across the routine and accident periods, likely because it lies far from all nuclear power stations. 

This shift in attention was temporary. Grouping prefectures by their distance from Fukushima (Figure~\ref{fig:decay-rate}), we found that search interest fell to roughly half its post-accident peak within about a week and to around 20\% within five weeks, at a similar rate in all three prefecture groups. Proximity to the 1F accident site thus shaped how intensely people searched but not how long they kept searching. Within a year, search interest in every region had returned close to its pre-accident level, leaving no lasting rise in attention. Finally, by 2012--2014, the correlation with distance had weakened and was no longer significant (Spearman's $\rho = -0.14$, $p = 0.356$; Supplementary Information, Figure S1b).

\subsubsection{Recovery and restart period (2012--2024): Reversion to localized attention}
\begin{figure}[htbp]
    \centering
    \includegraphics[width=\linewidth]{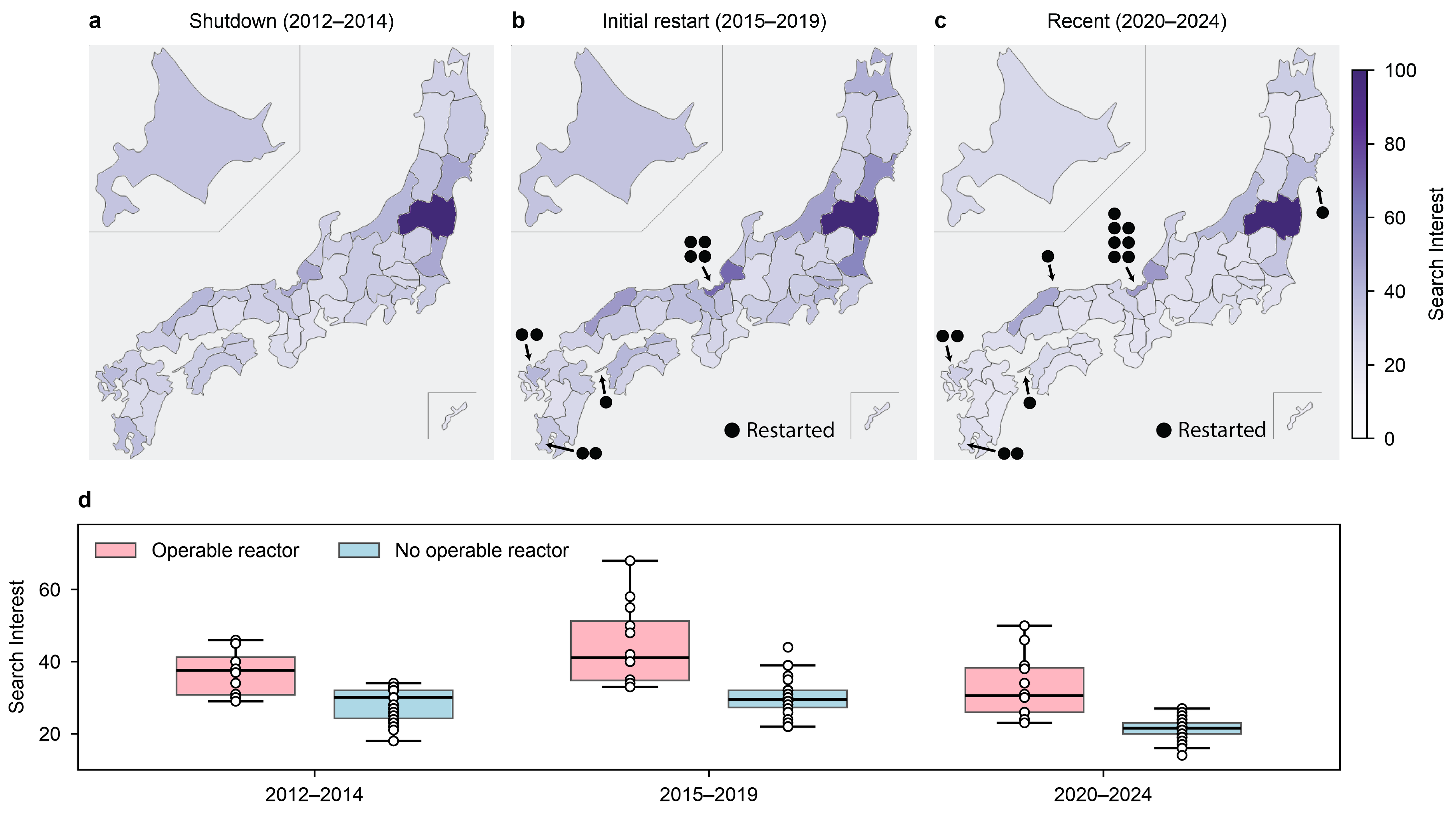}
    {\phantomsubcaption\label{fig:post-2011-shutdown}}
    {\phantomsubcaption\label{fig:post-2011-restart}}
    {\phantomsubcaption\label{fig:post-2011-recent}}
    {\phantomsubcaption\label{fig:post-2011-violin}}
    \caption{\textbf{Search Interest on the topic of "nuclear power plant" for the three post-2011 periods.} (a--c) Prefecture-level maps for the shutdown (2012--2014), initial restart (2015--2019), and recent (2020--2024) periods, on a shared color scale. Black dots in (b) and (c) mark the reactors that had restarted by the end of each period (9 by 2019 and 14 by 2024). (d) Search Interest in prefectures with and without an operable reactor in each period, excluding Fukushima Prefecture, which had the maximum of 100 in every period.}
    \label{fig:after_fukushima_summary}
\end{figure}

Figure~\ref{fig:post-2011-shutdown}--\subref{fig:post-2011-recent} maps search interest for the three periods after attention had returned close to its pre-accident level at the end of 2011: shutdown (2012--2014), initial restart (2015--2019, during which 9 reactors restarted), and recent (2020--2024, when 5 more restarted). Fukushima Prefecture still had the highest search interest throughout 2012--2024, likely reflecting sustained local concern after the accident. Elsewhere, the multi-nodal host-locality pattern observed during 2004--2010 re-emerged from 2012 and strengthened after domestic reactors began restarting in 2015. Excluding Fukushima, host prefectures again had significantly higher search interest than non-hosts in all three periods (Figure~\ref{fig:post-2011-violin}; Mann--Whitney U tests, all $p < 0.001$; Supplementary Information, Table S2). The probability of higher interest also rose from 0.857 (2012--2014) to 0.925 (2015--2019) and 0.956 (2020--2024), so the two groups became steadily easier to tell apart. This distinction likely reflects intensified local information-seeking in host prefectures, where restart decisions returned nuclear policy to local agendas after the shutdown period of 2012--2014.

\subsection{Content Analysis of Search Queries}

The analysis so far, has shown where and how strongly the public sought information related to nuclear power, and how this shifted after the 1F accident. It remains unclear, however, what exactly people searched for and how this varied by place, which matters most when deciding what information to provide during an accident (RQ3). To address this, we examined the most-searched queries of March 2011 along two dimensions: explanatory queries, through which the public sought to understand the unfolding event, and action-oriented queries, through which it sought guidance on how to respond. The full set of related queries for 11--31 March 2011 is provided in the Supplementary Information.

\subsubsection{Explanatory queries}

Because explanatory queries cover a broad range, we specifically captured searches for background information and for real-time updates. We did this by selecting queries that contain one of two distinct Japanese auxiliary terms, \mbox{\begin{CJK}{UTF8}{ipxm}とは\end{CJK}} ("what is") or \begin{CJK}{UTF8}{ipxm}情報\end{CJK} ("information"). Throughout this paper, we refer to them as "definition queries" and "update queries," respectively. Representative queries containing each term are shown in Table~\ref{tab:what-is-queries}. These queries confirm the difference in nuance between the two terms, with the former asking for definitions and the latter for current updates. They also fall into largely different categories, with "definition queries" centered on nuclear energy, radiation, and disaster mechanisms, and "update queries" centered on electricity supply, transportation, and the disaster situation. Queries in these categories remained prominent through the end of March 2011, but they were not static, as existing topics grew in depth and new topics appeared. Among "definition queries," those on nuclear energy and radiation became more specific, with searches such as "what is a sievert" and "what is re-criticality" appearing a few days after the 1F accident. Among "update queries," gasoline availability emerged as a new focus after the initial three days, with queries such as "gasoline information" rising sharply in the following week.

\begin{table}[h]
\small
\centering
\caption{Representative explanatory queries following the Fukushima 1F accident.}
\label{tab:what-is-queries}
\begin{tabular}{p{4cm} p{3cm} p{5cm}}
\toprule
\textbf{Query Type} & \textbf{Category} & \textbf{Representative Query} \\
\midrule
\multirow{3}{=}{\parbox{4cm}{\raggedright Definition queries\\ \begin{CJK}{UTF8}{ipxm}とは\end{CJK} ("what is")}} 
 & Nuclear Energy & "What is nuclear power?" \\
 & Radiation & "What is cesium?" \\
 & Natural Disaster & "What is soil liquefaction?" \\
\midrule
\multirow{3}{=}{\parbox{4cm}{\raggedright Update queries\\ \begin{CJK}{UTF8}{ipxm}情報\end{CJK} ("information")}} 
 & Electricity & "Planned blackout information" \\
 & Natural Disaster & "Safety information" \\
 & Transportation & "Highway information" \\
\bottomrule
\end{tabular}
\end{table}

\begin{figure}[htbp]
    \centering
    \includegraphics[width=\linewidth]{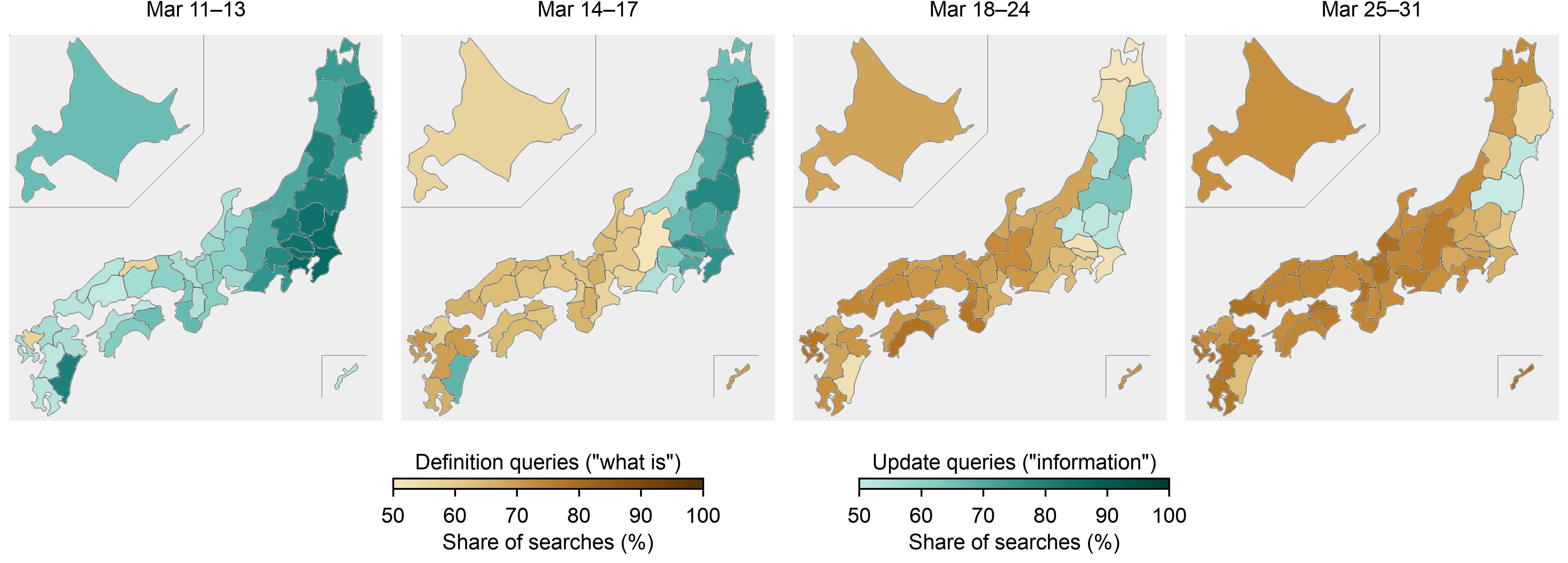}
    \caption{\textbf{Weekly search interest maps during March 2011 following the 1F accident.} Each prefecture is colored by the query type with the higher relative Google Trends Search Interest score, "definition queries" (\begin{CJK}{UTF8}{ipxm}とは\end{CJK}, "what is") in brown or "update queries" (\begin{CJK}{UTF8}{ipxm}情報\end{CJK}, "information") in teal.}
    \label{fig:toha_interest_maps}
\end{figure}

Figure~\ref{fig:toha_interest_maps} then compares, for each prefecture, the shares of "definition queries" and "update queries" in four windows between 11 and 31 March 2011. In the first three days after the earthquake and the 1F accident, "update queries" (teal) dominated nationwide, and the most-searched of them concerned the earthquake, safety information, transportation, and blackouts. Over the following two weeks, their dominance narrowed to the Tohoku and Kanto regions, and by late March it had contracted further to Miyagi and Fukushima, likely reflecting the prolonged disruptions there. However, because the earthquake and the nuclear accident occurred together, this heightened demand for real-time information reflects their combined effect and cannot be attributed to either event alone. Nevertheless, the maps show that the disaster gave real-time information demand a clear spatial pattern, centered on the affected regions, and that this pattern shifted markedly over time as it contracted toward the most affected prefectures.

Conversely, as the dominance of "update queries" contracted toward Fukushima, "definition queries" took the lead in prefectures farther away. In mid-March, the most-searched "definition queries" asked about basic terms, such as "what is a nuclear power plant," "what is radiation," and "what is a meltdown." Unlike "update queries," which spread across topics such as transportation, electricity, and gasoline, "definition queries" concentrated on nuclear energy and radiation. This suggests that, once the first wave of demand for real-time updates had passed, the public turned to understanding the nuclear accident itself, and that the accident likely raised awareness of nuclear topics. As time progressed, everyday queries such as "what is a smartphone" then began making up a growing share of the most-searched "definition queries," coinciding with the nationwide decline in search interest in "nuclear power plant" (Figure~\ref{fig:decay-rate}). 

\subsubsection{Action-oriented queries}

Beyond trying to understand the unfolding event, people also sought guidance on how to act. We traced this dimension in the same way, through queries containing one of two distinct Japanese auxiliary terms, \begin{CJK}{UTF8}{ipxm}方法\end{CJK} ("method") or \mbox{\begin{CJK}{UTF8}{ipxm}対策\end{CJK}} ("countermeasure"), which we refer to as "how-to queries" and "countermeasure queries," respectively. The two terms differ in nuance, with \begin{CJK}{UTF8}{ipxm}対策\end{CJK} being the more formal and generally referring to larger-scale measures. Accordingly, "how-to queries" asked how to carry out specific, often personal, tasks, such as "how to restore the gas meter," whereas "countermeasure queries" asked for measures against large-scale hazards and disruptions, such as "radiation leak countermeasures" (Table~\ref{tab:how-to-queries}).

\begin{table}[tbph]
\small
\centering
\caption{Representative action-oriented queries following the Fukushima 1F accident.}
\label{tab:how-to-queries}
\begin{tabular}{p{4cm} p{3cm} p{6cm}}
\toprule
\textbf{Query Type} & \textbf{Category} & \textbf{Representative Query} \\
\midrule
\multirow{3}{=}{\parbox{4cm}{\raggedright How-to queries\\ \begin{CJK}{UTF8}{ipxm}方法\end{CJK} ("method")}}
 & Gas & "How to restore the gas meter" \\
 & Radiation & "How to decontaminate after radiation exposure" \\
 & Other & "How to donate" \\
\midrule
\multirow{3}{=}{\parbox{4cm}{\raggedright Countermeasure queries\\ \begin{CJK}{UTF8}{ipxm}対策\end{CJK} ("countermeasure")}}
 & Electricity & "Planned blackout countermeasures" \\
 & Radiation & "Radiation leak countermeasures" \\
 & Natural Disaster & "Disaster preparedness measures" \\
\bottomrule
\end{tabular}
\end{table}

\begin{figure}[tbph]
    \centering
    \includegraphics[width=0.49\linewidth]{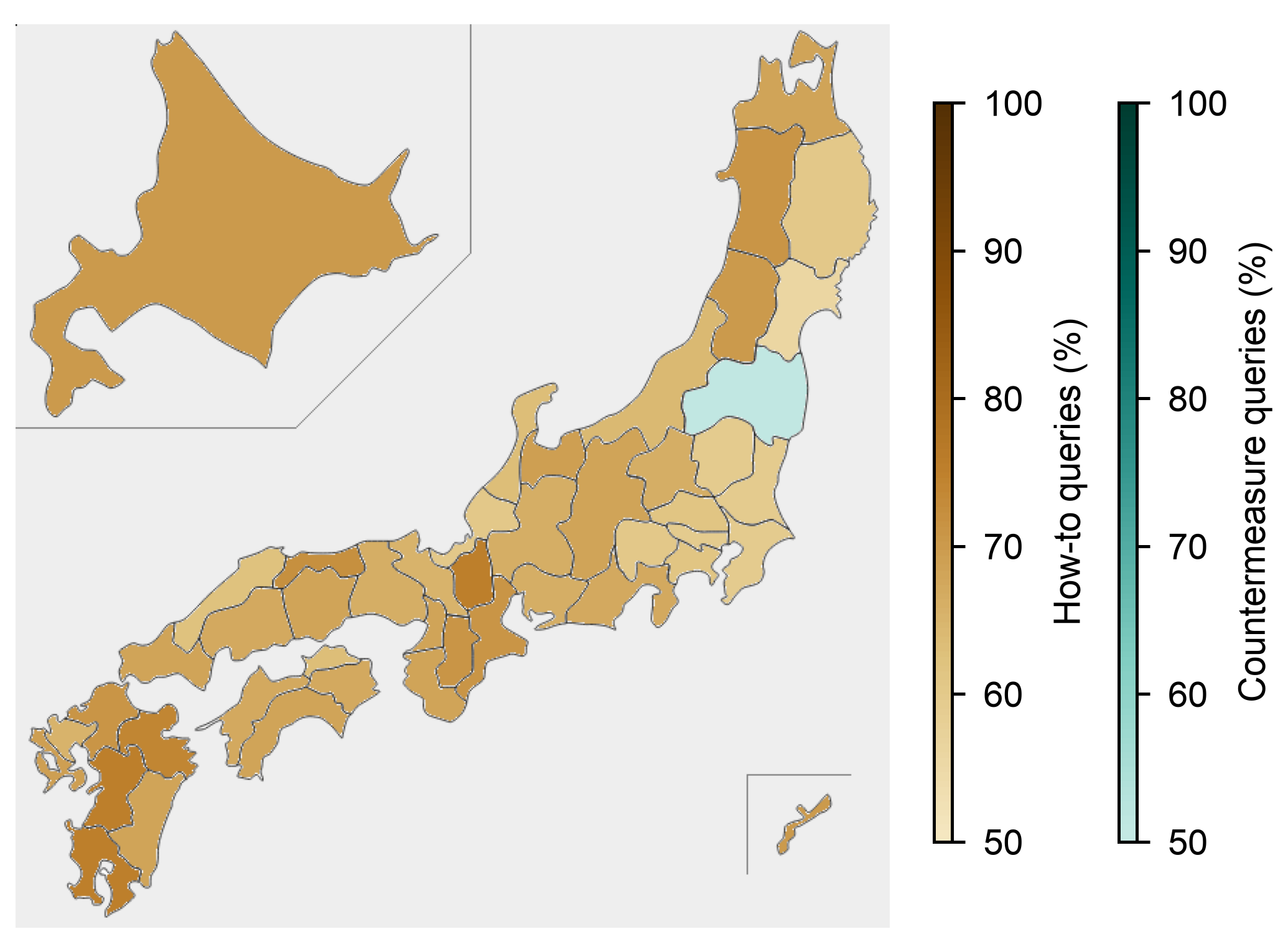}
    \caption{\textbf{Search interest map for action-oriented queries during 11–17 March 2011.} Each prefecture is colored by the query type with the higher relative Google Trends Search Interest score, "how-to queries" (\begin{CJK}{UTF8}{ipxm}方法\end{CJK}, "method") in brown or "countermeasure queries" (\begin{CJK}{UTF8}{ipxm}対策\end{CJK}, "countermeasure") in teal.}
    \label{fig:howto-interest-maps}
\end{figure}

We analyzed action-oriented queries only for the first week (11--17 March 2011), because in later weeks the top action-oriented queries included many everyday searches unrelated to the disaster, such as "diet methods" among the "how-to queries." Within that week, the spatial pattern was less distinct than for the explanatory queries. We attribute this to the nature of the two pairs, as the explanatory terms ask for different kinds of information, whereas both action-oriented terms ask for guidance, differ mainly in tone, and covered many of the same themes, such as radiation, electricity, and disasters (Supplementary Information, Figures S10 and S11). Even so, Fukushima was the only prefecture where "countermeasure queries" had the larger share, albeit by a small margin, whereas the more personal "how-to queries" were most prominent in the other prefectures (Figure~\ref{fig:howto-interest-maps}). Given the nuance of the two terms, this is not unexpected, as people near the accident were hit hardest, which likely shaped how they framed their searches.

\section{Discussion}
Effective public engagement and continued support around nuclear power go beyond capturing public attitudes and require understanding what citizens actually want to know, which surveys rarely measure. Through our data-driven analysis of two decades of Google Trends search data in Japan (2004–2024), we comprehensively mapped out, for the first time, how online information demand around nuclear power varies in space and time. The results allow us to address each of the three research questions in turn.

During routine periods, attention is structured by host-locality, with residents in prefectures hosting operational or planned reactors searching at substantially higher rates than residents elsewhere, and spillover to neighboring prefectures is generally small (RQ1). While this elevated local attention resembles classic NIMBY ("not in my backyard") effects \cite{dear1992understanding}, search signals are broader and may reflect interest from supporters and opponents alike, both of whom engagement efforts must reach. Where spillover does occur, it appears to result from physical proximity to a plant rather than a shared border alone. Tottori is a clear example. It has no operating or planned reactors of its own, yet it had the highest search interest among all such prefectures. This is likely because part of it lies within 30 km of the Shimane plant, a proximity that brought local involvement in the plant's affairs and, with it, attention across the border. This conditional spillover means that governments and stakeholders should clearly define the audience for public engagement on siting new reactors or waste-disposal facilities, and on regulatory hearings, by who is close enough to be affected, even across prefectural borders. Beyond Japan, the same approach can be adapted to map routine information demand and support public engagement in newcomer countries, such as the United Arab Emirates and Bangladesh, and in established nuclear nations debating fleet renewal or expansion, such as France and South Korea.

The Fukushima 1F accident then shows how the routine host-locality pattern gave way to a proximity-driven pattern in which attention declined with distance from Fukushima (RQ2). During 2011, attention concentrated near the area of impact, with search interest in Fukushima far exceeding that elsewhere. The post-accident surge, however, faded at a similar rate everywhere, so proximity shaped how intensely people searched rather than how long the surge lasted. The type of information sought also varied with proximity (RQ3). During March 2011, searches near Fukushima favored live updates, whereas those farther away turned increasingly to background information. Among action-oriented queries, Fukushima alone leaned toward more formal countermeasure guidance, while personal-scale how-to questions led everywhere else. Because the 1F accident was at its core a geographically anchored emergency, these proximity-driven patterns may also appear in similar emergencies outside the nuclear field and could inform communication strategies there. After 2011, although Fukushima kept the highest search interest in Japan, likely reflecting sustained local concern, the proximity-driven pattern did not last, and the host-locality pattern re-emerged and strengthened as restart decisions returned nuclear policy to local agendas. Consequently, host-prefecture communication channels are again becoming more important as Japan brings reactors back online, and they should be designed for the continuously heightened information demand we observed during routine periods. 

The content of the most-searched queries during March 2011 also suggests that much of the post-accident information demand was foundational rather than technical. The most-searched "definition queries" contained the most basic vocabulary of the domain, such as "what is a nuclear power plant" and "what is a meltdown," and the action-oriented queries on nuclear topics likewise named protective actions only in broad terms, such as "radioactivity countermeasures." These patterns suggest that for many in the searching public, the 1F accident may have been the first occasion on which the technical vocabulary of nuclear and radiation topics had to be actually understood rather than merely recognized. Without sufficient baseline knowledge of nuclear power operations or radiation risks, one therefore cannot easily formulate either queries that track the nuclear situation in real time or specific procedural queries. Between the two query types, the much lower overall search volume of action-oriented queries may also reflect a perception that individuals can do little about a reactor accident. Based on this empirical evidence, emergency communication following a major nuclear event should not assume high levels of technical literacy, but should actively anticipate and push information about hazards and protective actions that the public would not know to seek on their own. In addition, the searched queries grew more specific as the month progressed, with searches such as "what is a sievert" and "what is re-criticality" appearing a few days after the accident. This trajectory tracked how the public came to confront unfamiliar technical vocabulary as the situation unfolded, and public outreach during an accident should keep pace with this deepening demand.

Beyond communication strategies, spontaneous search behavior during an actual emergency also provides a basis for designing educational efforts. Nuclear energy is important yet hard to understand, and its accidents are rare but high-stakes, so maintaining widespread baseline literacy is inherently difficult. It is nevertheless important to establish a consistent education channel, for example by preparing accessible foundational content (what radiation is, what general protective actions exist) during routine periods. Rather than aiming for comprehensive information coverage, the effort should go into moving people beyond a vague sense of having heard a term to an actual understanding of the few concepts that matter most. This approach to education may also apply beyond nuclear energy to fields with complex concepts that are seldom needed by the public but can become critical during emergencies.

\section{Methods}
\subsection{Google Trends data}
Google Trends provides aggregate measures of search activity through its "Search Interest" metric, defined as the relative frequency of searches for a specified term or topic against all searches in a given time window and geography, normalized to a maximum of 100 within that window. Google defines a "topic" as a group of closely related queries, including their cross-language variants, which makes topics language-agnostic. We used the topic "\begin{CJK}{UTF8}{ipxm}原子力発電所\end{CJK}" (nuclear power plant) as the indicator of public attention to nuclear energy, chosen after comparing its relative search intensity with alternatives such as "nuclear energy" and "nuclear power."

We collected two types of data from Google Trends. The first is Search Interest for all 47 prefectures in five non-overlapping windows: pre-accident (2004--2010), accident (2011), shutdown (2012--2014), initial restart (2015--2019), and recent (2020--2024). Within each window, values are normalized to the prefecture with the highest Search Interest. We also retrieved weekly Search Interest for each prefecture in 2011, normalized to that prefecture's own 2011 peak, to resolve the post-accident decay. The second type addresses how the content of searches varied during March 2011. For this, we retrieved prefecture-level comparison-mode data for two pairs of Japanese auxiliary terms, an explanatory pair and an action-oriented pair. Within each selected time period, comparison mode returns each prefecture's share of search interest between the two paired terms. In addition, we manually reviewed the top related queries for each auxiliary term to identify those directly related to the disaster.

\subsection{Geo-location and nuclear status}
We measured each prefecture's distance from Fukushima Prefecture as a proxy for its distance from the Fukushima 1F accident site. This was done by calculating the GRS80 geodesic distance between the two prefectural government offices, with the office coordinates obtained from the Geospatial Information Authority of Japan. Historical reactor operating status was taken from data published by the Japan Atomic Industrial Forum.

\subsection{Statistical analysis}
Two types of statistical analysis were performed. First, non-parametric two-sided Mann–Whitney U tests were used to compare Search Interest between prefecture groups within a period. The test's null hypothesis is that the two groups share the same Search Interest distribution, so that a prefecture drawn at random from one group is equally likely to rank above or below one drawn from the other. The null is rejected when the two-sided $p$-value falls below the significance level, set at $\alpha = 0.05$. For each test we also report the common-language effect size, which is the share of all comparison--baseline prefecture pairs in which the comparison prefecture has the higher Search Interest, with ties counted as one half. We refer to this as the probability of higher interest. Second, two-sided Spearman's rank correlation tests were performed to evaluate the relationship between Search Interest and distance from Fukushima Prefecture, with $\alpha = 0.05$.

\section{Data availability}
The data that support the findings of this study are available at \url{https://github.com/KU-Yifan/Google-Trends-JP-Nuclear}.

\section{Code availability}
The code that supports the findings of this study is available at \url{https://github.com/KU-Yifan/Google-Trends-JP-Nuclear}.

\section{Author contributions}
Y.S. conceived the study, collected the data, performed the analysis, and wrote the manuscript. K.K. and B.C.M. supervised the study and revised the manuscript. All authors read and approved the final version.

\section{Competing interests}
The authors declare no competing interests.

\bibliographystyle{unsrtnat}
\bibliography{references} 

\end{document}